\documentclass[prl,superscriptaddress,twocolumn,10pt,nofootinbib,aps, floatfix]{revtex4-1}
\usepackage{amssymb}
\usepackage{amsmath}
\usepackage{appendix}
\usepackage{times}
\usepackage{graphicx}
\usepackage{subeqnarray}

\usepackage{color}

\begin{document}

\title {Floquet-Engineered Odd-Parity Altermagnetic Higher-Order Topology in a Two-Dimensional Antiferromagnet Cr$_2$CH$_2$}
	
\author{Xiaorong Zou}
\affiliation{Center for 2D Quantum Heterostructures (2DQH), Institute for Basic Science (IBS), Suwon 16419, Republic of Korea}

\author{Hyeon Suk Shin}
\affiliation{Center for 2D Quantum Heterostructures (2DQH), Institute for Basic Science (IBS), Suwon 16419, Republic of Korea}
\affiliation{Department of Chemistry, Sungkyunkwan University, Seobu-ro 2066, Suwon 16419, Republic of Korea}
\affiliation{Department of Energy, Sungkyunkwan University, Seobu-ro 2066, Suwon 16419, Republic of Korea}

\author{Baibiao Huang}
\affiliation {School of Physics, State Key Laboratory of Crystal Materials, Shandong University, Jinan 250100, China}

\author{Yanmei Zang}
\affiliation{Department of Energy, Sungkyunkwan University, Seobu-ro 2066, Suwon 16419, Republic of Korea}
\affiliation{Department of Energy Science, Sungkyunkwan University, Seobu-ro 2066, Suwon 16419, Republic of Korea}

\author{Ying Dai}
\email{daiy60@sdu.edu.cn}
\affiliation {School of Physics, State Key Laboratory of Crystal Materials, Shandong University, Jinan 250100, China}

\author{Chengwang Niu}
\email{c.niu@sdu.edu.cn}
\affiliation {School of Physics, State Key Laboratory of Crystal Materials, Shandong University, Jinan 250100, China}

\author{Chang-Jong Kang}
\email{cjkang87@cnu.ac.kr}
\affiliation{Department of Physics, Chungnam National University, Daejeon 34134, Republic of Korea}
\affiliation{Institute for Sciences of the Universe, Chungnam National University, Daejeon 34134, Republic of Korea}

\author{Chang Woo Myung}
\email{cwmyung@skku.edu}
\affiliation{Center for 2D Quantum Heterostructures (2DQH), Institute for Basic Science (IBS), Suwon 16419, Republic of Korea}
\affiliation{Department of Energy, Sungkyunkwan University, Seobu-ro 2066, Suwon 16419, Republic of Korea}
\affiliation{Department of Energy Science, Sungkyunkwan University, Seobu-ro 2066, Suwon 16419, Republic of Korea}
\affiliation{Department of Quantum Information Engineering, Sungkyunkwan University, Seobu-ro 2066, Suwon, 16419, Korea}
	
\begin{abstract}
Periodic driving provides a platform to dynamically tailor quantum states of matter, yet its impact on symmetry-protected topological phases remains incompletely understood. Here, we demonstrate that periodic driving enables the realization of an odd-parity altermagnetic (AM) higher-order topological insulator (HOTI) phase in the Cr$_2$CH$_2$ monolayer. In equilibrium, Cr$_2$CH$_2$ is a 2D antiferromagnetic (AFM) HOTI protected by $\mathcal{C}_3$ rotational symmetry, characterized by a symmetry indicator $\chi^{(3)}$ = $\{-2,1\}$ and robust corner states. Under circularly polarized light (CPL), the system develops a f-wave altermagnetic state governed by the symmetry $[C_{2}||\overline{3}_{001}]$ with odd-parity spin splitting. Despite substantial Floquet-induced band renormalization, the $\mathcal{C}_3$-protected corner states remain intact over a broad range of driving strengths, highlighting the altermagnetic higher-order topology under Floquet driving. As the light intensity increases, the system gradually evolves into an altermagnetic semimetallic state. These results establish a direct connection between magnetism and topology in a periodically driven AFM system, offering a route toward the control of coupled spin and topological transport.
\end{abstract}
	
\maketitle
\date{\today}

Antiferromagnets have recently emerged as a fertile platform for exploring novel quantum states, owing to their vanishing net magnetization, robustness against external magnetic fields, and potential for ultrafast spin dynamics~\cite{AFM1,AFM2,Baltz2018,2018AFMSA}. Beyond hosting a variety of nontrivial topological phases, such as topological semimetals and insulators with symmetry-protected boundary states~\cite{Tang2016,mnbiteex1,Shao2019prl,Niu2020PRL,2023NCAFMTI,2024PRLAFMSM}, they provide a unique setting in which topology and magnetism are intrinsically intertwined~\cite{Nakatsuji2015,AFM3,2024NMAFM}. Altermagnets, a recently proposed magnetic paradigm, broadens the conventional picture of antiferromagnets by enabling symmetry-protected spin splitting in systems with fully compensated moments~\cite{2022PRXalter3,2022PRXalter,2024AFMalterfeng,2025NRPAM}. Such states typically exhibit even-parity spin textures in momentum space (e.g., $d$-, $g$-, and $i$-wave channels). More recently, this framework has been extended to incorporate odd-parity spin splitting (e.g., $p$- and $f$-wave channels), giving rise to a distinct class of odd-parity magnetic phases~\cite{2022PRXalter2,2024naturealter,2024SAalter,2026PRLAMODD,2026PRLAMODD2,2026OHEAM}. As a rapidly evolving research direction, altermagnetism has expanded the range of candidate materials~\cite{2023PRLmagnon,2024prlmntealter,2024NCalter,2024naturemntealter,2024NLalter} and enables diverse spintronic functionalities, including spin transport and magnetoelectric control~\cite{2021PRLSPIN,2022NESPIN,2024PRLalterspin}. However, existing studies of altermagnetic topological phases are predominantly restricted to even-parity channels~\cite{2024AMTIPRB,2024PRBAMHOTI,2026AFMAM}, with odd-parity states remaining largely unexplored.

Floquet engineering provides a framework for controlling the electronic structure and magnetism of quantum materials~\cite{2013PRLflo,2015APflo,2022ARflo}. Through time-periodic driving, band structures can be effectively reconstructed and symmetries selectively modified without altering the underlying chemical composition~\cite{2020NRPflo,2023PSSflo}. In antiferromagnetic systems, Floquet driving can break time-reversal symmetry while preserving crystalline symmetries, thereby inducing odd-parity spin splitting, manifested as p-wave- and f-wave altermagnetic spin textures with well-defined angular dependence~\cite{2026OHEAM,2026PRLAMODD,2026PRLAMODD2}. Concurrently, Floquet-induced band renormalization can reshape the topological properties of the system~\cite{2016PRLflo,2024floNL,2024PRBflo1,2024PRBflo}. Notably, periodic driving has been demonstrated experimentally to induce anomalous Hall responses in graphene and realize Floquet-Bloch bands in materials such as black phosphorus and the surface states of Bi$_2$Se$_3$~\cite{2013SCIflo,2020NPflo,2023natureflo}. These results highlight that, under periodic driving, symmetry breaking, spin-texture reconstruction, and topological evolution are intrinsically coupled, offering a unified route to control magnetic and topological responses within a single platform.

Higher-order topological insulators (HOTIs), as an extension of topological phases, host boundary states localized at hinges or corners and are protected by crystalline symmetries~\cite{highorderfirst,highorderrenyafei,2021NRPn1,2021NRPn12}. Although such phases have been realized in various magnetic systems~\cite{higherordereuln2as2,highorderbieuo,2023AFMHOTI,2023nanoletterlirunhan}, their realization in intrinsic 2D antiferromagnetic materials remains relatively limited. Here, we demonstrate that the two-dimensional antiferromagnet Cr$_2$CH$_2$ monolayer realizes a higher-order topological insulating phase. The system exhibits in-gap floating edge states together with corner states in a triangular nanoflake, protected by $\mathcal{C}_3$ rotational symmetry and consistent with the symmetry indicator $\chi^{(3)}$ = $\{-2,1\}$. These bulk, boundary, and corner signatures collectively establish its higher-order topology. Under circularly polarized light, the system breaks the time-reversal-related symmetry $[C_{2}T||E]$ while preserving $[C_{2}||P]$ and crystalline symmetry $[E||C_3]$, resulting in an odd-parity spin splitting with an $f$-wave angular dependence, characteristic of an altermagnetic spin texture. Over a broad range of driving strengths, the corner states remain robust and retain spatial distributions consistent with the crystalline symmetry. This robustness arises from the selective symmetry action of periodic driving. With increasing driving strength, the bulk gap is progressively reduced and eventually closes, leading to an altermagnetic semimetallic state.

First-principles calculations were carried out within the framework of density functional theory (DFT) using the Vienna \textit{ab initio} simulation package (VASP)~\cite{Kresse,Kresse1}. The exchange-correlation effect was described by the Perdew--Burke--Ernzerhof (PBE) functional within the generalized gradient approximation (GGA)~\cite{pbevasp}. A plane-wave basis set with a kinetic energy cutoff of 550~eV was used. The Brillouin zone was sampled by a $13 \times 13 \times 1$ $\Gamma$-centered $k$-point mesh. To eliminate spurious interactions between periodic images, a vacuum layer of 20~\AA\ was introduced along the out-of-plane direction. The atomic positions and lattice constants were fully relaxed until the Hellmann--Feynman forces on each atom were smaller than $0.01~\mathrm{eV/\AA}$, with an energy convergence criterion of $10^{-6}$~eV. To account for the on-site Coulomb interaction of Cr-$3d$ electrons, the GGA+U scheme was applied with $U = 3$ eV. Maximally localized Wannier functions were constructed using the WANNIER90 package~\cite{wannier90}.

\begin{figure} 
	\centering
	\includegraphics[width=1\linewidth]{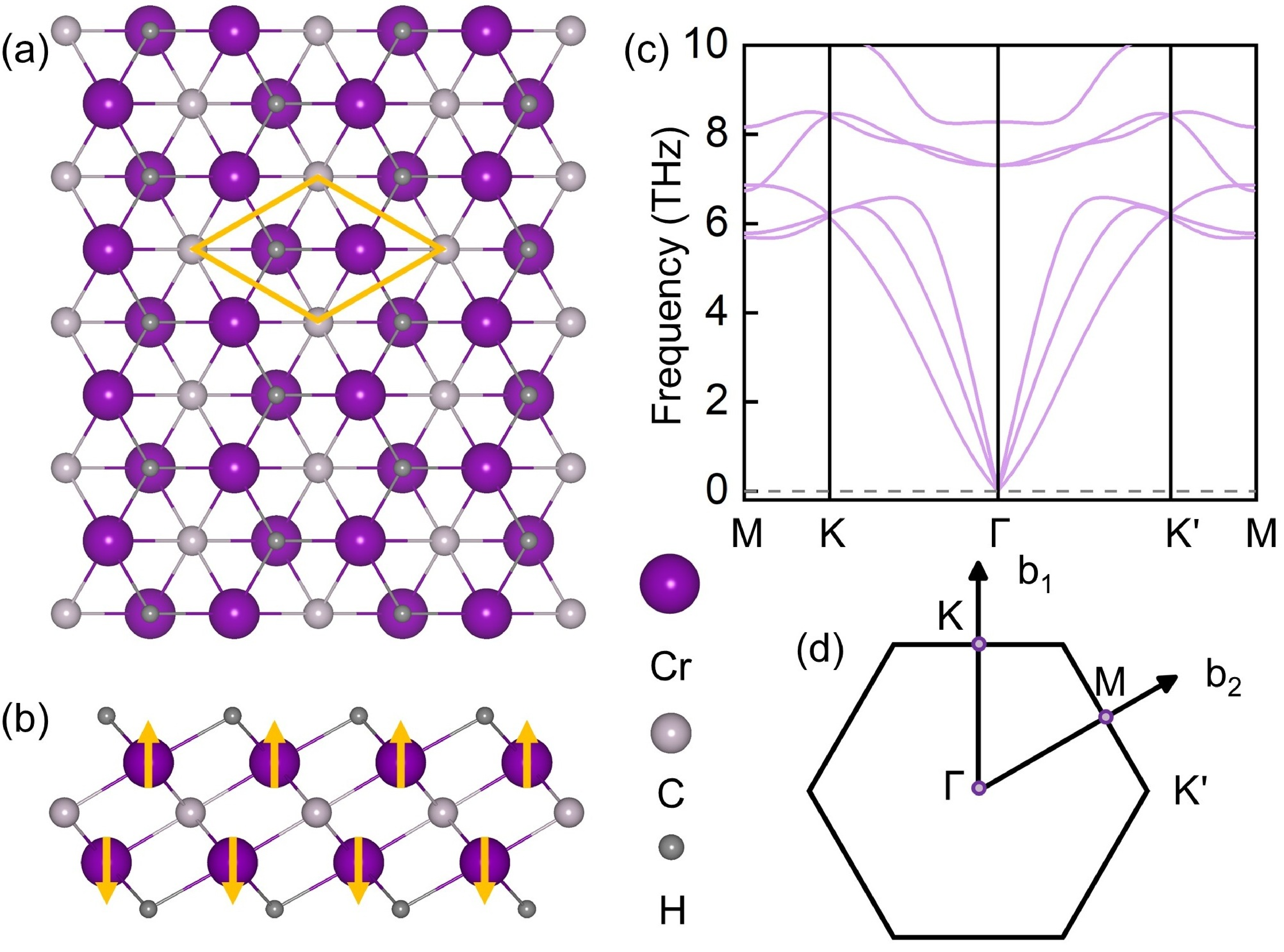}
	\caption{(a) Top view and (b) side view of the two-dimensional Cr$_2$CH$_2$ monolayer. The unit cell is shown in yellow. (c) Phonon spectrum of the Cr$_2$CH$_2$ monolayer. The absence of imaginary frequencies confirms its dynamical stability. (d) Schematic illustration of the hexagonal Brillouin zone with high-symmetry points. The reciprocal lattice vectors are denoted by $b_1$ and $b_2$.}
	\label{structure}
\end{figure}

\begin{figure}
	\centering
	\includegraphics[width=1\linewidth]{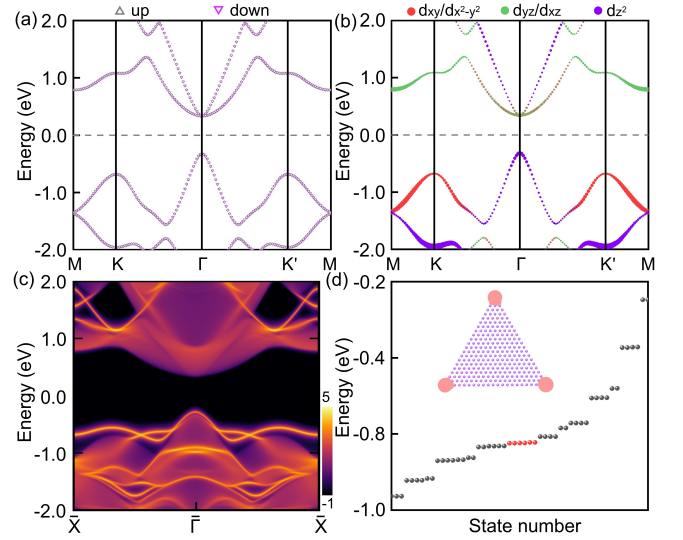}
	\caption{(a) Spin-resolved band structure of the Cr$_2$CH$_2$ monolayer, showing spin degeneracy enforced by $[C_{2}||P]$ and $[C_{2}T||E]$. (b) Orbital-resolved band structure of the Cr$_2$CH$_2$ monolayer. (c) Edge states and (d) discrete energy spectra of a triangular nanoflake for the 2D AFM Cr$_2$CH$_2$ monolayer. The Fermi level $E_F$ is set to be zero. The pink dots indicate the in-gap corner states. Insets show the corresponding real-space charge distributions of the six degenerate corner states, as marked by red dots.} 
	\label{afmhoti}
\end{figure}

We first examine the magnetic ground state of the two-dimensional Cr$_2$CH$_2$ monolayer. By comparing the total energies of different magnetic configurations,
we find that the antiferromagnetic (AFM) state presented in Fig.~\ref{structure}(b) is energetically favored over the ferromagnetic (FM) state by approximately 0.76 eV, establishing a stable two-dimensional A-type AFM ground state consistent with previous report~\cite{2025PRBc2ch}. The absence of imaginary phonon spectrum further confirms its dynamical stability, as shown in Fig.~\ref{structure}(c). Based on this AFM configuration, we analyze the system within the framework of spin-group symmetry. The symmetry can be described by a set of spin–space operations, including $[C_{2}||P]$, $[C_{2}T||E]$, and $[E||C_3]$, which collectively impose constraints on both spin and momentum degrees of freedom. In particular, $[C_{2}||P]$ combines a twofold spin rotation with spatial inversion, while $[C_{2}T||E]$ denotes a combined symmetry of time reversal and spin rotation that maps states between opposite momenta in the collinear AFM state. The operation $[E||C_3]$ corresponds to the threefold rotational symmetry of the lattice, constraining the angular structure in momentum space. Under these symmetry constraints, the energy eigenvalues satisfy $\varepsilon_s(\boldsymbol{k})=\varepsilon_{-s}(-\boldsymbol{k})$ and $\varepsilon_s(\boldsymbol{k})=\varepsilon_s(-\boldsymbol{k})$, which further lead to $\varepsilon_s(\boldsymbol{k})=\varepsilon_{-s}(\boldsymbol{k})$. As a result, spin degeneracy is enforced throughout the Brillouin zone as illustrated in Fig.~\ref{afmhoti}(a). This symmetry-enforced degeneracy provides the foundation for the subsequent emergence of nontrivial spin textures. 

\begin{figure*}
	\centering
	\includegraphics[width=0.6\linewidth]{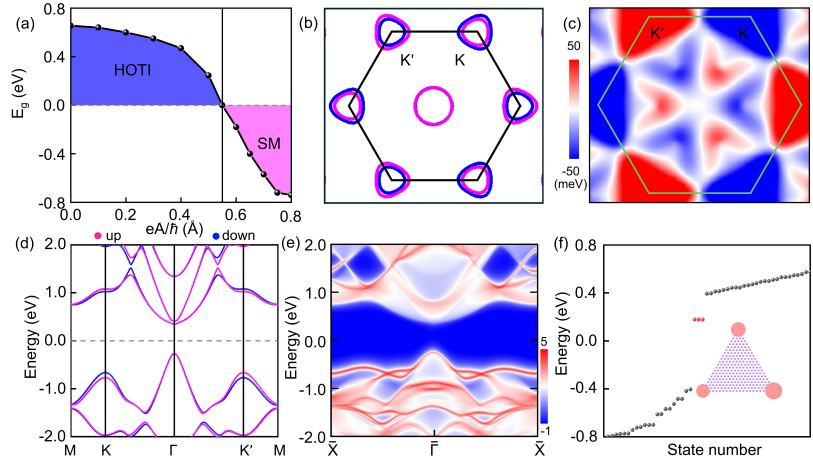}
	\caption{
		(a) Evolution of the bulk gap of Cr$_2$CH$_2$ monolayer as a function of light intensity at $\hbar\omega=9$ eV. (b) Constant energy map at $E - E_F = -0.8$ eV and (c) momentum-resolved spin splitting between the highest spin-up and spin-down valence bands under CPL with $\hbar\omega=9$ eV and $eA/\hbar=0.2~\mathrm{\AA}^{-1}$, exhibiting a characteristic $f$-wave altermagnetic pattern. (d) Spin-resolved band structure at $\hbar\omega=9$ eV and $eA/\hbar=0.2~\mathrm{\AA}^{-1}$, revealing odd-parity spin splitting. (e) Edge states and (f) energy discrete spectra of a triangular nanoflake for Cr$_2$CH$_2$ monolayer at $\hbar\omega=9$ eV and $eA/\hbar=0.2~\mathrm{\AA}^{-1}$. The Fermi level $E_F$ is set to be zero. The pink dots indicate the in-gap corner states. The insets show the corresponding real-space charge density distributions of the three degenerate corner states, marked by red dots.}
	\label{amhoti}
\end{figure*}

We then examine the electronic structure and identify a bulk energy gap of about 0.655 eV near the Fermi level, indicating an insulating phase. To probe the boundary properties, we compute the edge spectrum using a Green’s function approach. Within the bulk gap, the floating edge states are clearly resolved in Fig.~\ref{afmhoti}(c); however, these states remain gapped and do not form dispersive channels connecting the valence and conduction bands. The absence of gapless edge states, together with a fully insulating bulk, distinguishes the system from conventional first-order topological insulators and instead indicates a higher-order topological phase. To further substantiate the higher-order topology, we construct a triangular nanoflake. The calculated energy spectrum reveals a set of in-gap states near the Fermi level, as displayed in Fig.~\ref{afmhoti}(d). Analysis of their real-space charge density distribution shows that these states are strongly localized at the three corners of the nanoflake and exhibit identical profiles at symmetry-equivalent positions, a hallmark of two-dimensional higher-order topological insulators. 

In $\mathcal{C}_3$-symmetric systems, the topology is characterized by the symmetry indicator $\chi^{(3)}$ and the associated fractional corner charge $Q_{c}^{(3)}$~\cite{highorderinvariants,c3invariant,higher2020invariant}. The occupied bands are classified by the rotation eigenvalues $e^{i\frac{2\pi}{3}(p-1)}$ ($p=1,2,3$), and the symmetry-resolved indices are defined as
\begin{equation}
	[K_{p}^{(3)}] = \#K_{p}^{(3)} - \#\Gamma_{p}^{(3)},
\end{equation}
where $\#K_{p}^{(3)}$ and $\#\Gamma_{p}^{(3)}$ denote the number of occupied states with eigenvalue $e^{i\frac{2\pi}{3}(p-1)}$ at the $K$ and $\Gamma$ points, respectively. Therefore, the $\chi^{(3)}$ and $Q_{c}^{(3)}$ are given by
\begin{align}
	\chi^{(3)} &= ([K_{1}^{(3)}],\, [K_{2}^{(3)}]), \\
	Q_{c}^{(3)} &= \frac{2e}{3}([K_{1}^{(3)}] + [K_{2}^{(3)}]) \ \mathrm{mod}\ 2e.
\end{align}
For the present system, we obtain $\chi^{(3)} = (-2, 1)$, corresponding to a fractional corner charge $Q_{c}^{(3)} = 4e/3$. This result is consistent with the real-space localization of the corner states, confirming the higher-order topological character of the system.

To incorporate the effect of periodic driving, we consider a time-dependent tight-binding Hamiltonian,
\begin{equation}
	H(\boldsymbol{k},t)=
	\sum_{m,n,j}
	t_j^{mn}(t)\,
	e^{i\boldsymbol{k}\cdot\boldsymbol{R}_j}
	c_m^{\dagger}(\boldsymbol{k})
	c_n(\boldsymbol{k}) .
\end{equation}
Here, $t_j^{mn}(t)$ denotes the time-dependent hopping amplitude between orbital $m$ in the home cell and orbital $n$ in the $j$th cell. For normally incident circularly polarized light, the vector potential is taken as
\begin{equation}
	\mathbf{A}(t)=A(\sin\omega t,\cos\omega t,0),
\end{equation}
and the hopping amplitudes are modified through the Peierls substitution,
\begin{equation}
	t_j^{mn}(t)=
	t_j^{mn}
	\exp\left[
	i\frac{e}{\hbar}
	\mathbf{A}(t)\cdot\boldsymbol{d}_j^{mn}
	\right],
\end{equation}
where $\boldsymbol{d}_j^{mn}$ denotes the displacement vector between orbital $m$ in the home cell and orbital $n$ in the $j$th cell. In the high-frequency off-resonant regime, where $\hbar\omega$ is chosen to be larger than the bandwidth to avoid crossings between Floquet bands, the driven system can be described by the effective Floquet Hamiltonian
\begin{equation}
	H_{\mathrm{eff}}
	=
	H_0+
	\frac{1}{\hbar\omega}
	\sum_{n\geq1}
	\frac{[H_{-n},H_{+n}]}{n}
	+\mathcal{O}(\omega^{-2}),
\end{equation}
where $H_{\pm n}$ denote the Fourier components in the frequency space. The commutator term originates from virtual photon absorption and emission processes, acting as an off-resonant correction that renormalizes the band structure in momentum space. From a symmetry perspective, periodic driving acts selectively on the spin-group symmetries. In equilibrium, the spin-group constraints enforce $\varepsilon_s(\boldsymbol{k})=\varepsilon_{-s}(\boldsymbol{k})$ leading to spin degeneracy throughout the Brillouin zone. Under circularly polarized light, the time-reversal-related symmetry $[C_2T||E]$ is broken, whereas $[C_2||P]$ and the crystalline $C_3$ rotational symmetry are preserved. Consequently, the system still obeys $\varepsilon_s(\boldsymbol{k})=\varepsilon_{-s}(-\boldsymbol{k})$, but no longer satisfies $\varepsilon_s(\boldsymbol{k})=\varepsilon_s(-\boldsymbol{k})$. The spin degeneracy is therefore lifted, giving rise to a momentum-odd spin splitting,
\begin{align}
	\Delta E(\boldsymbol{k})
	&=
	\varepsilon_{s}(\boldsymbol{k})
	-
	\varepsilon_{-s}(\boldsymbol{k}), \\
	\Delta E(\boldsymbol{k})
	&=
	-\Delta E(-\boldsymbol{k}) .
\end{align}

We next elucidate the evolution of band topology and momentum-space spin structure under periodic driving at a fixed frequency $\hbar\omega = 9~\mathrm{eV}$. By tracking the bulk gap as a function of the light intensity $eA/\hbar$, we find that the system remains a higher-order topological insulating phase over a broad regime $eA/\hbar \in [0,\,0.55]~\text{\AA}^{-1}$, as illustrated in Fig.~\ref{amhoti}(a). To illustrate this regime, we focus on a representative case at $eA/\hbar = 0.2~\text{\AA}^{-1}$. The edge spectrum exhibits in-gap floating edge states that do not connect the valence and conduction bands, while a triangular nanoflake hosts corner states with charge density distribution concentrated at three symmetry-related corners, shown in Figs.~\ref{amhoti}(e) and (f), providing direct evidence of higher-order topology. 

Having verified the higher-order topology, we turn to the spin texture under periodic driving. As shown in Fig.~\ref{amhoti}(d), the spin-resolved bands exhibit a pronounced odd-parity spin splitting, with opposite spin polarizations at the $K$ and $K'$ points, characteristic of an altermagnetic state. The constant energy map at $E - E_F = -0.8$ eV in Fig.~\ref{amhoti}(b) exhibits a threefold symmetric angular dependence, indicative of an $f$-wave spin texture in momentum space. This behavior can be understood from symmetry perspective: the combined spin-space symmetry $[C_{2}||P]$ and $[E||C_3]$ generate an effective symmetry $[C_{2}||\overline{3}_{001}]$, which constrains the spin texture in momentum space and enforces a threefold angular modulation. Consistently, the momentum-resolved gap distribution between the highest spin-up and spin-down valence bands in Fig.~\ref{amhoti}(c) exhibits the same modulation, providing an independent signature of the $f$-wave character. Taken together, these results demonstrate that the system realizes an odd-parity altermagnetic higher-order topological insulating phase.

To further assess the robustness of our results, we vary the driving frequency within the high-frequency regime while fixing the amplitude at $eA/\hbar \approx 0.2~\text{\AA}^{-1}$. The corner states persist throughout, remaining sharply localized at the three symmetry-equivalent corners of the nanoflake without noticeable degradation. This behavior indicates that the periodically driven altermagnetic higher-order topological phase is governed primarily by crystalline $C_3$ rotational symmetry, rather than by the specific choice of driving frequency, as shown in Fig. S1.

\begin{figure}
	\centering
	\includegraphics[width=1\linewidth]{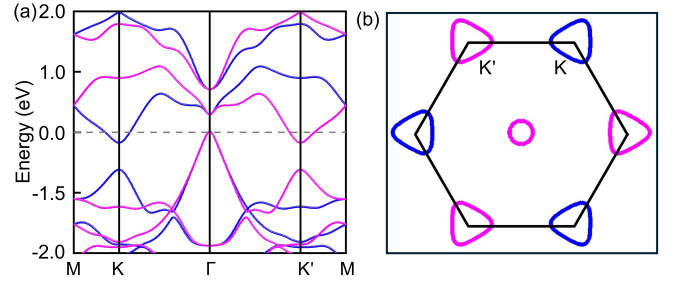}
	\caption{(a) Band structure with odd-parity spin split at $\hbar\omega = 9~\mathrm{eV}$ and $eA/\hbar = 0.6~\text{\AA}^{-1}$. (b) Constant energy map at $E - E_F = -0.1$ eV, exhibiting a characteristic $f$-wave altermagnetic pattern.} 
	\label{tsm}
\end{figure}

With further increasing driving strength, the bulk gap is progressively suppressed and closes at $eA/\hbar \approx 0.55~\text{\AA}^{-1}$, beyond which the system enters a semimetallic phase. Notably, the characteristic $f$-wave spin texture persists in this regime, as confirmed by both the band structure and the constant energy map at $E - E_F = -0.1$ eV, as displayed in Figs.~\ref{tsm}(a) and (b). 

In this work, we investigate the impact of periodic driving on the magnetic structure and higher-order topology of the two-dimensional antiferromagnet Cr$_2$CH$_2$. We first establish that Cr$_2$CH$_2$ monolayer realizes a $C_3$-protected higher-order topological insulating phase characterized by robust corner states. Under circularly polarized light (CPL), the momentum-space spin structure is reconstructed, driving a transition from an antiferromagnetic state to an altermagnetic phase with odd-parity spin splitting. Importantly, the crystalline symmetries that protect the corner states remain intact, allowing the higher-order topology to persist over a broad range of driving strengths. As the driving strength is further increased, the system evolves into an altermagnetic semimetallic state. These results demonstrate that periodic driving enables the coexistence of odd-parity altermagnetism and higher-order topology within a single system. More broadly, our findings reveal an intrinsic connection among symmetry, magnetism, and topology, and establish periodic driving as a viable route for engineering coupled magnetic and topological phenomena in quantum materials.

This work was supported by Institute for Basic Science (IBS-R036-D1) and the Taishan Scholar Program of Shandong Province. We are grateful for the computational support from the Korea Institute of Science and Technology Information (KISTI) (KSC-2023-CRE-0355, KSC-2023-CRE-0261, KSC-2023-CRE-0502, KSC-2024-CRE-0144, KSC-2024-CRE-0088, KSC-2024-CRE-0117). Computational work for this research was partially performed on the Olaf supercomputer supported by IBS Research Solution Center and GPU cluster supported by Ministry of Science and ICT (MSIT) and the National IT Industry Promotion Agency (NIPA).

\end{document}